\documentclass[aps,pra,preprint, superscriptaddress,amsmath,amssymp,showpacs]{revtex4-1}
\usepackage{natbib}
\usepackage[T1]{fontenc}
\usepackage[latin1]{inputenc}
\usepackage[usenames,dvipsnames]{color}
\usepackage{rotating}
\usepackage{bm}        % \boldsymbol (bold math)
\usepackage{amssymb}   % \mathbb     (blackboard bold)
\usepackage{amsmath} 
\usepackage{subfig}
\usepackage{physics}

\usepackage{caption}

\usepackage{bm}
\usepackage{CJK}
\def\jcp#1#2#3{J.~Chem.~Phys.~{\bf #1},\ #2\ (#3)}

\def\pra#1#2#3{Phys.~Rev.~A~{\bf #1},\ #2\ (#3)}
\def\prl#1#2#3{Phys.~Rev.~Lett.~{\bf #1},\ #2\ (#3)}

\def\k1{k_1}
\def\k2{k_2}
\def\q1{q_1}
\def\q2{q_2}

\def\({\left (}
\def\){\right )}
\def\[{\left [}
\def\]{\right ]}

\newcommand{\beq}{\begin{equation}}
\newcommand{\eeq}{\end{equation}}

\begin{document}
\date{\today}
\flushbottom \draft
\title{Steady-state Fano coherences in a V-type system driven by polarized incoherent light} 

\author{Suyesh Koyu}
\affiliation{Department of Physics, University of Nevada, Reno, NV 89557, USA}
\author{Amro Dodin}
\affiliation{Department of Chemistry, MIT, Cambridge, Massachusetts  02139}
%\author{Amar Vutha}
%\affiliation{Department of Physics, University of Toronto, Toronto, Ontario, M5S 3H6, Canada}
\author{Paul Brumer}
\affiliation{Chemical Physics Theory Group, Department of Chemistry, and Center for Quantum Information and Quantum Control, University of Toronto, Toronto, Ontario, M5S 3H6, Canada}
\author{Timur V. Tscherbul}
\affiliation{Department of Physics, University of Nevada, Reno, NV 89557, USA}\email[]{ttscherbul@unr.edu}

\begin{abstract}

We explore the properties of steady-state Fano coherences generated in a three-level V-system continuously pumped by polarized incoherent light in the absence of coherent driving. 
By solving the nonsecular Bloch-Redfield quantum master equation we obtain the ratio of the stationary coherences to excited-state populations, $\mathcal{C} = (1+\frac{\Delta^2}{\gamma(r+\gamma)} )^{-1}$, which quantifies the impact of steady-state coherences on excited-state dynamical observables of the V-system. The ratio is maximized when the excited-state splitting $\Delta$ is small compared to either the  spontaneous decay rate $\gamma$ or the incoherent pumping rate $r$. We demonstrate that while the detrimental effects of a strongly decohering environment generally suppress the coherence-to-population ratio by the factor $\simeq \gamma_d/\gamma$, an intriguing regime exists where the $\mathcal{C}$ ratio displays a maximum as a function of  the dephasing rate  $\gamma_d$.  We attribute the surprising dephasing-induced enhancement of stationary Fano coherences to the environmental suppression   of destructive interference of individual incoherent excitations generated at different times.
We further clarify the physical basis for the steady-state Fano coherence, whose imaginary part is identified with the non-equilibrium flux across a pair of qubits coupled to two independent thermal  baths,  unraveling a direct connection between the seemingly unrelated phenomena of incoherent driving of multilevel quantum systems and non-equilibrium quantum transport in qubit networks.
The real part of the steady-state Fano coherence is found to be proportional to the deviation of excited-state populations from their values in thermodynamic equilibrium, making it possible to observe signatures of steady-state Fano coherences in excited-state populations.  Finally, we put forward an experimental proposal for observing  steady-state Fano coherences by detecting the total fluorescence signal emitted by Calcium atoms excited by polarized vs. isotropic incoherent light. Our analysis paves the way toward further theoretical and experimental studies of non-equilibrium coherent steady states in thermally driven atomic and molecular systems, and for the exploration of their  potential role in biological processes.

%FOR INTRO: We explore the properties of the steady-state noise-induced Fano coherences that arise  in a three-level V-system  irradiated by polarized incoherent light.
%FOR CONCLUSIONS: we show that the Bloch-Redfield master equations for the V-system driven by polarized incoherent light are equivalent to those describing the transport problem involving two coupled two-level systems interacting with a hot and a cold bath. In this picture, 

%Our results suggest the presence of new, experimentally observable   regimes of stationary quantum coherent dynamics, in atomic and molecular systems driven by polarized incoherent light.
%with implications for the design of quantum heat engines and photovoltaic devices.
%We discuss the prospects for observing the long-lived noise-induced coherences in strongly driven Rydberg atoms.
\end{abstract}

% and require nearly perfect alignment of  the transition dipole moments of the V-system under isotropic  incoherent excitation
%oscillate in the underdamped regime ($\Delta/\gamma > \bar{n}$) and

\maketitle
\clearpage
\newpage

\section{introduction}

Quantum coherence is widely regarded as an essential resource \cite{Streltsov:17} for quantum information processing \cite{Ladd:10}, quantum sensing  \cite{Degen:17}, and quantum interferometry \cite{Cronin:09}. The inevitable interaction of quantum systems with an external environment is generally believed to lead to the decay of quantum coherence in a process known as decoherence \cite{Schlosshauer:97}. As decoherence upsets the unitary time evolution necessary for the successful operation of quantum bits and sensors, understanding and controlling decoherence mechanisms is a key research goal of quantum science and technology.

An additional motivation to study decoherence comes from recent experimental and theoretical studies  of quantum effects in biological processes, such as photosynthetic energy transfer \cite{Cheng:09,Pachon:12,Brumer:18}, vision \cite{Tscherbul:15a,Dodin:19},  and avian magnetoreception \cite{Hiscock:16}. These studies have indicated that non-trivial quantum effects such as coherence and entanglement \cite{Cheng:09,Pachon:12} can persist under noisy conditions typical of biological environments at room temperature. In all of the experimental studies performed to date, quantum coherence was introduced into the system by means of coherent (ultrashort) laser pulses \cite{Cheng:09,Pachon:12,Brumer:18}. However, biologically relevant photosynthetic light-harvesting is driven by natural sunlight, which is an incoherent mixture of number states,  raising an important question: Can excitation by incoherent light alone generate coherence in multilevel quantum systems \cite{Brumer:12,Tscherbul:14,Tscherbul:15,Cao:14,Dodin:16,Dodin:16a,Brumer:18}? 

% \cite{Cheng:09,Pachon:10,Brumer:18}

Surprisingly, the answer to this question is ``yes'', although the coherences are distinctly different from coherent light-induced effects \cite{Dodin:19}. Incoherent radiative transitions between manifolds of closely spaced energy levels can interfere to produce {\it noise-induced Fano coherences} in multilevel atomic and molecular systems even in the absence of coherent driving \cite{Agarwal:70,Fleushhauer:92,Hegerfeldt:93,Kozlov:06,Tscherbul:14,Tscherbul:15,Dodin:16,Dodin:16a,Koyu:18}.
The interference manifests itself  through the cross terms in the quantum optical (Bloch-Redfield)  master equation, which correspond to the interaction of a common incoherent light mode with the  dipole moments of the transitions from the same initial state to different final states \cite{Kozlov:06,Tscherbul:14,Dodin:16,Dodin:16a,Koyu:18}.
%Quantum coherence can be generated in multilevel atomic and molecular systems through Fano interference, whereby

Initial theoretical studies of Fano coherences highlighted their significance in the context of lasing without inversion \cite{Fleushhauer:92} and quantum jumps in trapped ions \cite{Hegerfeldt:93}.  Agarwal and Menon \cite{Agarwal:01} explored the conditions under which the three-level V-system (see Fig. 1) pumped by  incoherent light  approaches  thermodynamic equilibrium in the long-time limit. They showed that in the absence of symmetry between incoherent pumping and spontaneous decay, non-trivial steady-state coherences can arise, leading to non-equilibrium steady states.  Kozlov, Rostovtsev, and Scully \cite{Kozlov:06}  showed that incoherent pumping of the V-system with degenerate upper levels can create  coherent population-locked states, which depend on the initial state of the system \cite{Kozlov:06}.  Recent theoretical work by Dorfman, Scully, Mukamel, and co-workers \cite{Scully:11,Dorfman:13} reinvigorated  interest in noise-induced Fano coherences by showing that they can enhance the efficiency of quantum heat engines by breaking the principle of detailed balance.  Note, however, that a  problem has recently been discovered \cite{Creatore:13} with the master equation approach used in Refs.~\cite{Scully:11,Dorfman:13} which gives negative population dynamics at long times.
%A recent theoretical study found enhanced  thermoelectric performance of molecular heat engines in the presence of coherence \cite{Chen:17}.

We have explored the dynamical properties of Fano coherences in realistic V-systems with nondegenerate excited states, establishing the existence of two dynamical regimes depending on the ratio of the excited-state splitting $\Delta$ to the radiative decay rate $\gamma$ \cite{Tscherbul:14,Dodin:16,Koyu:18}. In the underdamped regime characterized by $\Delta/\gamma\gg 1$, Fano coherence exhibit damped oscillations that decay on the timescale $\tau\simeq 1/\gamma$ \cite{Tscherbul:14,Dodin:16}. In the overdamped regime, where the excited-state splitting is smaller than the radiative decay rate, we observe long-lived quasisteady-state coherences with the lifetime $\tau=\frac{2}{\gamma}(\frac{\Delta}{\gamma})^{-2}$ \cite{Tscherbul:14,Dodin:16}. We have also explored the dynamics of the V-system driven by polarized incoherent light \cite{Dodin:18} and proposed an experimental scheme for observing noise-induced coherence dynamics in Calcium atoms driven by incoherent radiation.

% the  in  demonstrated that they can potentially be used to break detailed balance and thereby increase the efficiency of quantum heat engines such as photovoltaic cells and photodetectors 

The majority of theoretical studies of noise-induced Fano coherences have focused on the dynamics induced by {\it isotropic} incoherent light, whose polarization ($\bm{\epsilon}_\lambda$) and propagation ($\mathbf{k}$) vectors point randomly in all directions. The coherence properties of isotropic incoherent light [such as its first-order degree of coherence $g^{(1)}(\tau)$] are  independent of $\bm{\epsilon}_\lambda$  and $\mathbf{k}$ \cite{Mandel:95}. As a consequence,  the  light-matter interaction only depends on the angles between the  dipole moment vectors $\bm{\mu}_{ij}$ and $\bm{\mu}_{kl}$ for radiative transitions between the system eigenstates $|i\rangle \leftrightarrow |j\rangle$ and  $|k\rangle \leftrightarrow |l\rangle$  \cite{Tscherbul:15}. The resulting quantum master equation treats   incoherent excitation and decay dynamics on an equal footing, leading to the decay of quantum coherences in the long-time limit and the establishment of the canonical equilibrium steady state \cite{Agarwal:01}, although the time it takes to reach this state can be extremely long for nearly-degenerate levels \cite{Tscherbul:14,Dodin:16,Koyu:18}.

In contrast, when a quantum system is excited by polarized incoherent light, only one polarization mode of the light interacts with the transition dipole moment vectors \cite{Dodin:18,Dodin:19}, whereas spontaneous emission still occurs in all directions (in the absence of an external cavity).  As a result of the imbalance between incoherent pumping and spontaneous emission, the excitation with polarized incoherent light can lead to the emergence of nonequilibrium steady states (NESS) featuring  coherences  in the energy eigenstate basis and populations deviating significantly from their expected canonical values \cite{Agarwal:01}. %Importantly, these phenomena occur solely under incoherent driving, {\it i.e.}, in the absence of a second thermal bath. 
% which could create steady-state coherence through non-equilibrium transport.
We note that these steady states are distinct from the more commonly occurring NESS that arise due to the additional selection rules for excitation with polarized light. For instance, the $\Delta m_J=\pm 1$  electric dipole selection rule implies that the $|J=1,m_J=0\rangle$ excited state cannot be populated from the ground $|J=0,m_J=0\rangle$ state irradiated by incoherent light polarized in the $x$-direction \cite{Dodin:18}. While important for optical pumping, these NESS manifest themselves through the appearance of dark states, and do not feature coherences in the energy eigenstate basis of the system, so we will not consider them further in this work. 
% Thus, pumping with polarized incoherent light will create a non-equilibrium distribution of $|Jm\rangle$ excited states, a phenomenon that lies at the heart of optical pumping. We will not consider such ``trivial'' 

In contrast to the case of excitation with isotropic incoherent light, very little attention has been devoted to the NESS and steady-state Fano coherences induced  by polarized incoherent excitation of multilevel quantum systems. Agarwal and Menon  derived analytical expressions for steady-state Fano coherences in a three-level V-system driven by broadband incoherent radiation \cite{Agarwal:01}. We have shown that the steady-state coherences can occur in  Calcium atoms driven by polarized  incoherent light \cite{Dodin:18}. However, even the most basic properties of steady-state Fano coherences  (such as their dependence of the system and excitation parameters, and their sensitivity to decoherence)  remain unexplored, despite widespread occurrence of polarized sunlight in nature \cite{Cronin:11}, and the resultant possibility of  noise-induced coherences playing a role in biological excitation processes.
 
Here, we present a theoretical analysis of steady-state Fano coherences in a three-level V-system driven  by polarized incoherent excitation. The V-system serves as a minimal model, in which to study the quantum dynamics of noise-induced coherence generation \cite{Tscherbul:14,Cao:14} and energy transfer \cite{Kassal:14,Cao:14,Tscherbul:18}.  We obtain analytic results for the steady-state coherences and explore their dependence on the system parameters. Our results suggest that the coherences are  maximized when the excited-state splitting is small compared to the radiative decay rate ({\it i.e.}, in the large-molecule limit \cite{Tscherbul:14}). %The steady-state Fano coherences also tend to increase with  incoherent pumping intensity. 
We clarify the physical origin of the coherences by establishing an equivalence between the quantum master equation describing the V-system excited by polarized incoherent radiation and that describing the dynamics of non-equilibrium heat transport across a pair of quantum two-level systems (qubits) coupled to two independent thermal baths \cite{Li:15}. Finally, we present an experimental scenario for observing the  signatures of steady-state Fano coherences in the fluorescence signal emitted by calcium atoms excited by polarized incoherent light.
Our analysis paves the way toward further theoretical and experimental studies of NESS and steady-state Fano coherences in thermally driven atomic and molecular systems, and for the exploration of their  potential role in natural processes  induced and sustained by solar light.

%These noise-induced coherent dynamics may be relevant for many physical, chemical and biological processes driven by incoherent light such as visual phototransduction, nanoscale heat transfer, and the transfer of light energy during the initial steps of photosynthesis process, photo voltaic cells. Notably, the study of Fano coherences has been restricted to theoretical calculations. As such, the possibility for the experimental observation of Fano coherences in atomic and molecular systems remains an important open question

%and  made longer by increasing the pumping intensity, which could make the noise-induced  coherences easier to observe experimentally.

In the rest of this paper, we will present the details of our theoretical calculations based on the nonsecular Bloch-Redfield (BR) quantum master equation (Sec. II). In Sec. III we report the results for steady-state Fano coherences as a function of the system and excitation parameters. In Sec. IIIA we establish the physical basis for steady-state Fano coherences using a two-bath transport model. The effects of relaxation and decoherence are examined in Sec. IIIB. An experimental proposal for observing Fano coherences in Calcium atoms in put forward  in Sec. IV. Section V summarizes  and discusses the main results of this work.

\section{Theory and Methods: Analytical solutions to Bloch-Redfield equations in the steady state}

To model the quantum dynamics of the V-system driven by polarized incoherent radiation, we use the Bloch-Redfield (BR) quantum master equation for the system's density matrix  in the eigenstate basis \cite{BPbook,Agarwal:01,Tscherbul:14,Dodin:16,Dodin:16a}
\begin{align}
    \Dot{\rho}_{ii} &= -(r_{i}+\gamma_{i}+\Gamma_i) \rho_{ii} + r_{i} \rho_{gg} - \sqrt{r_{a}r_{b}} \rho^{R}_{ab} \nonumber
    \\ 
    \Dot{\rho}_{ab} &= -\frac{1}{2}(r_{a}+r_{b}+\gamma_{a}+\gamma_{b}+2\gamma_{d})\rho_{ab} -i\Delta \rho_{ab} + \sqrt{r_{a}r_{b}} \rho_{gg} - \frac{1}{2} \sqrt{r_{a}r_{b}} (\rho_{aa}+\rho_{bb}) \label{QME1} 
\end{align} 
where $\rho_{ab}=\rho^{R}_{ab}+i\rho^{I}_{ab}$ is the coherence between the excited eigenstates $\ket{a}$ and $\ket{b}$ [see Fig.~\ref{fig:1}(a)], $\gamma_i$ is the spontaneous decay rate of the $i$-th excited-state level,  $r_i = \bar{n} \gamma_i$ is the incoherent pumping rate, and  $\bar{n}=[1-e^{\hbar\omega_{ag}/{kT}}]^{-1}$ is the average photon occupation number of the incoherent radiation filed, which we assume to be polarized in the $x$-direction of the space-fixed coordinate frame \cite{Dodin:18}.  The environmental relaxation and decoherence channels (to be considered  in Sec. IIIB) are described by the phenomenological rates $\Gamma_i$ and $\gamma_d$.  Note that Eq. (\ref{QME1})  assumes that  non-radiative coupling with an external  environment ({\it e.g.} phonons) cannot generate coherence between the eigenstates of the V-system \cite{Benenti:17}, {\it i.e.}, that Fano coherence can only be generated by the light-matter interaction through the $ \sqrt{r_{a}r_{b}} \rho_{gg} $ term in Eq. (\ref{QME1}). Isotropic incoherent excitation in the presence of coherence-generating system-phonon interactions have been considered elsewhere \cite{Svidzinsky:11,Tscherbul:15,Tscherbul:18,Dodin:19}.

%We will assume that non-radiative relaxation is negligible ($\Gamma_i=0$) throughout the rest of this paper. Non-radiative decoherence effects ($\gamma_d>0$) will be 

We will consider the case of a symmetric V-system with $\gamma_{a}=\gamma_{b}=\gamma$, $r_{a}=r_{b}=r$, and $\Gamma_{a}=\Gamma_{b}=\Gamma$ which contains all of the essential physics while drastically simplifying the analytical solution of the BR equations (\ref{QME1}).  
Imposing the trace conservation condition for the ground-state population $\rho_{gg}= 1-2\rho_{aa}$, we obtain the BR equations for the symmetric V-system in the absence of environmental relaxation and decoehrence
\begin{align}
    \Dot{\rho}_{aa} &= -(3r+\gamma +\Gamma) \rho_{aa} - r \rho^{R}_{ab}+r  \nonumber \\
    \Dot{\rho}^{R}_{ab} &= -3r \rho_{aa} - (r+\gamma+\gamma_d)\rho^{R}_{ab} + \Delta \rho^{I}_{ab}+r \nonumber \\
    \Dot{\rho}^{I}_{ab} &= -\Delta \rho^{R}_{ab}-(r+\gamma+\gamma_d)\rho^{I}_{ab} \label{QME2} 
\end{align} 
Here, as above, the phenomenological  parameters $\Gamma$ and $\gamma_{d}$ account for the effects of relaxation and decoherence between the excited states of the V-system due to, {\it e.g.}, non-radiative processes. These effects will be explored in Sec. IIIB below.  In Sec. III, we will begin by considering only the radiative processes, setting $\Gamma=\gamma_{d}=0$.

The BR quantum master equations (\ref{QME2}) can be rewritten in matrix-vector notation as \cite{Dodin:16}
\begin{equation} \label{state_vector_DE}
    \Dot{\mathbf{x}}(t) = \mathbf{A}\mathbf{x}(t) + \mathbf{d},
\end{equation} 
where $\mathbf{x}(t)=[\rho_{aa}(t),\rho^{R}_{ab}(t),\rho^{I}_{ab}(t)]^{T}$ is the state vector in the Liouville representation \cite{Dodin:16}, $\mathbf{d}=[r,r,0]^{T}$ is the driving vector  corresponding to the V-system  being initially in the ground state [$\rho_{gg}(t=0)=1$], and the coefficient  matrix $\mathbf{A}$ is given by 
\begin{equation}
    \mathbf{A} = 
\begin{bmatrix}
    -(3r+\gamma) & -r & 0 \\
    -3r  & -(r+\gamma) & \Delta \\
    0 & -\Delta  &  -(r+\gamma) 
\end{bmatrix}
\end{equation}
In this work, we are primarily interested in the steady-state behavior of the V-system irradiated by incoherent light. Setting $\Dot{\mathbf{x}}(t)=0$ in Eq. (\ref{state_vector_DE}) we obtain the steady state solution
\begin{equation} \label{SSVec} 
    \mathbf{x}_{s} = - \mathbf{A}^{-1} \mathbf{d}
\end{equation} 
The steady-state solution is unique only if the coefficient matrix $\mathbf{A}$ is non-singular, {\it i.e.}, $\det(\mathbf{A}) =3r^{2}(r+\gamma) -(3r+\gamma)[\Delta^{2}+(r+\gamma)^{2}] \ne 0$. The determinant is non-zero in general but  it tends to zero as $\Delta \to 0$ and $r/\gamma\to \infty$, leading to initial condition-dependent population-locked steady states  similar to those explored in Ref. \cite{Kozlov:06}. However, due to, {\it e.g.}, differential level shifts experienced by the excited levels interacting with the pumping field, most real-world V-systems will have $\Delta > 0$. As a result,  $\det(\mathbf{A}) \neq 0$ and  the steady-state solution  (\ref{SSVec}) is uniquely defined.

\section{Steady-state Fano coherences: Results and Discussion }

Substituting the inverse of matrix $\mathbf{A}$  into Eq. (\ref{SSVec}), we find analytical expressions for excited-state populations and the real part of Fano coherence  in the steady state  \cite{KoyuThesis:19}
\begin{align}
    \rho_{aa} &= \frac{r[\Delta^{2}+(r+\gamma)^{2}-r(r+\gamma)]}{(3r+\gamma)[\Delta^{2}+(r+\gamma)^{2}]-3r^{2}(r+\gamma)} \nonumber \\
    \rho^{R}_{ab} &= \frac{r\gamma(r+\gamma)}{(3r+\gamma)[\Delta^{2}+(r+\gamma)^{2}]-3r^{2}(r+\gamma)}  \label{SS_Dyn} 
\end{align} 
with the imaginary part of the coherence  $\rho^{I}_{ab} = -\frac{\Delta}{(r+\gamma)} \rho^{R}_{ab}$, in agreement with the results of Ref.~\cite{Agarwal:01}. Below we analyze the remarkable properties of  steady-state Fano coherences  not explored in Ref.~\cite{Agarwal:01}, and study  their dependence on  the excited-state splitting $\Delta/\gamma$ and the incoherent pumping rate $r=\bar{n}\gamma$.  The effects of relaxation and decoherence are considered in the following section. 
%conditions under which are maximized as a function of
% as well as  effects to Sec. as well as of  decoherence  who did not analyze them further. Such an analysis is present below.

First, we observe that neither the real nor imaginary part of the Fano coherence vanishes in the steady state, in contrast to the previously explored cases of incoherent driving with isotropic (unpolarized) light \cite{Tscherbul:14,Dodin:16,Koyu:18}. Figure~ \ref{fig:2}  shows the time evolution of excited-state populations and coherences in the V-system driven by polarized incoherent light. The results are obtained by numerical integration of Eq. (\ref{QME2}).
%We observe that the populations and coherences reach their steady-state values, which are in excellent agreement with
The equilibration timescale depends on the dynamical regimes of the V-system classified in our previous work \cite{Dodin:18,KoyuThesis:19}, which are determined by the value of excited-state splitting $\Delta/\gamma$ and the incoherent pumping rate $r=\bar{n}\gamma$.  Regardless of the transient dynamics,  the V-system eventually reaches a steady state characterized by nonzero values of  Fano coherences shown by the dashed lines in Fig.~ \ref{fig:2}, which are in excellent agreement with the analytical expressions (\ref{SS_Dyn}).

It may seem surprising that incoherent driving  excites a coherent superposition rather than an incoherent mixture of energy eigenstates.
The presence of coherences seems incongruent with an incoherent nature of the light source, which is thought of as having an entirely random phase at every instant.
For such a source excitations generated at different times will pick up a different phase from the fluctuating light field, yielding a random phase that vanishes upon averaging over the realizations of the incoherent source.
However, this phase averaging argument does not hold for excitations generated at the same instant.
In this case, both excitations will accrue the same (randomly selected) phase from the incoherent drive, resulting in an in-phase superposition of the two eigenstates.
That is, while the relative phase between excitations generated at different times is uncorrelated, the relative phase of simultaneous excitations is deterministic since both arise by interacting with the same state of the light field \cite{Dodin:19}.
%[Detailed description in "Light-induced processes in nature: Coherences in the establishment of the nonequilibrium steady state in model retinal isomerization" A. Dodin, and P. Brumer; J. Chem. Phys. 150 (18); p. 18430 (2019).]

%The emergence of a coherent steady-state, and indeed of the previously reported quasi-stationary coherences  \cite{Tscherbul:14,Dodin:16,Dodin:18} is unexpected given the randomly varying phase of incoherent light.
To further understand the counterintuitive emergence of the coherent steady-state, we  consider the dynamics of the contributing coherent excitations.
As  noted  earlier in this section, an incoherent light source generates a coherent drive due to the presence of simultaneous excitations from the ground state to two different excited states.
This process generates an in-phase superposition of two energy eigenstates.
Once this excitation is generated, it undergoes unitary evolution, accruing a periodic relative phase with frequency $\Delta$.
Moreover, it will slowly dephase with rate $\gamma_d$ and relax to the ground state with rate $\gamma + r+\Gamma$ [see Eq.~(\ref{QME1})].
The timescales and effects of these contributions on the dynamics of a single excitation are shown schematically in  Fig.~\ref{fig:timescales_dephasing}(a).
%Fig. (timescales.png).
%[See attached file for the timescales schematic and the associated .svg source file.]

In incoherent driving, the source is assumed to be continuously acting on the system, generating new excitations at every instant in time. At each instant, therefore, a new in-phase superposition is generated and undergoes the dynamics described above. The resulting ensemble dynamics are then obtained by summing over the contributions of all of these different excitations, as shown in Fig.~\ref{fig:timescales_dephasing}(b).%(ensemble_dephasing.png).
If the system frequency $\Delta$ is much larger than the excitation decay rate ($\sim \gamma$), then excitations generated at earlier times will have accrued some phase due to unitary evolution in the excited state manifold while new excitations will always produce an in-phase superposition.
Summing over these contributions then leads to an integral over different phases, producing an ensemble dephasing with excitations generated at different times carrying different phase that eventually averages to near zero.
However, when the frequency is much slower than the excitation lifetime, nearly all excitations will decay before accruing an appreciable phase.
As a result, nearly all excitations will be in the initially prepared in-phase superposition, leading to the previously reported quasi-stationary coherences \cite{Tscherbul:14,Dodin:16}.
Generally, the interplay of these different processes need not result in an incoherent steady state, as new coherent superposition may be refreshed quickly enough to maintain coherence in the long-time limit.

In the case of unpolarized excitation that obeys the fluctuation-dissipation constraint $r = \bar{n} \gamma$, the rate of generating new excitations $r$ and their decay rate $\sim$$\gamma$ are directly related.
This bound ensures that the delicate interplay of generating new excitations, their coherent evolution and subsequent decay are balanced to yield an incoherent thermal steady-state.
However, if these rates are allowed to vary independently, it becomes possible to generate residual coherences as the balance between these different processes is broken.
As discussed in the framework of the thermal transport model in Sec. IIIA below, polarized driving  breaks this constraint by involving field modes at different temperatures, the high temperature ${x}$-polarization and the low-temperature isotropic vacuum. In this case, the rates are no longer balanced resulting in a coherent NESS.

The second important observation apparent from Eq. (\ref{SS_Dyn}) and Fig.~\ref{fig:2} is that the values of excited-state populations in the steady state deviate from those expected in canonical thermodynamic equilibrium 
\begin{equation} \label{CanEqn_NoC}
    \rho^\text{(c)}_{aa} = \frac{r}{(3r+\gamma)}
\end{equation}
  The deviation of excited-state populations from the Boltzmann distribution then follows from Eq. (\ref{SS_Dyn}) 
 \begin{align}
    \rho^{R}_{ab} &= \frac{\Delta \rho_{aa}}{\rho^\text{(c)}_{aa}} \label{CanEqn} \\
    \rho^{I}_{ab} &= -\frac{\Delta}{(r+\gamma)} \rho^{R}_{ab} \label{CanEqn_Imag}
\end{align}
 These expressions are a central result of this work. They establish that  polarized incoherent driving leads to the formation of a coherent non-equilibrium steady-state (NESS) characterized by substantial Fano coherences in the energy eigenstate basis of the V-system.

  Significantly, Eq. (\ref{CanEqn}) suggests that the relative deviation of excited-state populations in the NESS from their values in canonical thermodynamic equilibrium  (\ref{CanEqn_NoC}) is equal to the real part of the Fano coherence.  
  In other words, \textit{the real part of the steady-state coherence is directly proportional to the difference in the excited state populations with and without Fano coherence}.
 %Further, the imaginary part of the coherence is proportional to the real part of the coherence in the steady state.
    Equation (\ref{CanEqn})  thus provides a direct connection between steady-state Fano coherences and excited-state populations  of the V-system continuously driven by a polarized incoherent radiation field. 
  As such,  Eq.~(\ref{CanEqn}) enables  direct experimental detection of the steady-state Fano coherences through measuring the deviation of excited-state populations  from their canonical equilibrium values (\ref{CanEqn_NoC}), as shown in Sec. IV.  Because it is typically much easier to detect eigenstate populations than coherences, steady-state Fano coherences may be significantly easier to observe experimentally than their transient counterparts \cite{Dodin:18}.
 
 %In the following section, we will explore the dependence of steady-state coherences on the V-system parameters and excitation conditions. 
 %CONCLUSIONS: Correlates with the presence of non-zero steady-state Fano coherences in the system via remarkably simple relations, which
 % to  find the regimes under which  the steady-state coherences is maximized.

% This expression suggests that in the steady state, the relative deviation of the excited state populations from their canonical equilibrium values (\ref{CanEqn_NoC}) is equal to the real part of the Fano coherence. Further, the imaginary part of the coherence is proportional to the real part of the coherence in the steady state. In other words, \textit{the real part of the steady-state coherence is directly proportional to the difference in the excited state populations with and without Fano coherence}. This is a significant observation for two reasons. First, it provides a direct connection between Fano coherences and the excited-state populations in the steady state. Second, it is usually much easier to detect eigenstate populations than coherences. As shown below, Eq.~(\ref{CanEqn}) enables the direct experimental detection of the Fano coherence through measuring the excited state populations in the steady state. \par 

The physical origin of the non-equilibrium coherences (to be further clarified in Sec. IIIA below) can be traced back to Fano interference between the electric dipole transitions induced by the interaction with a common incoherent light field mode.
%   \textcolor{OrangeRed}{Comment on polarization modes, the importance of x-polarized incoherent light, and fluctuation-dissipation relations}.
    This interference is manifested in the population-to-coherence coupling terms in the BR equation (\ref{QME2}), which show that the presence of a positive real part of the coherence $\rho^{R}_{ab}(t)$, suppresses the excited state populations. \par

%\subsection{Steady-state coherences in the absence of non-radiative relaxation and decoherence}

We now turn to the study of the dependence of steady-state Fano coherences on the dimensionless parameters $\bar{n}$ and $\Delta / \gamma$ that control the dynamics of the incoherently driven V-system in the absence of relaxation and decoherence.  
Figure \ref{fig:3}(a) shows the dependence of  steady-state Fano coherences on the average occupation number of thermal photons $\bar{n}=r/\gamma$ in the pumping field. The coherences  increase gradually with the pumping intensity and approach a common limit that does not depend on $\Delta/\gamma$.
As shown in Fig.~\ref{fig:3}(b)  the steady-state coherence are nearly independent of  the excited-state splitting $\Delta/\gamma$ in the weak pumping limit  $\bar{n}\ll 1$. At higher pumping intensities, the coherences begin to decline monotonously with inclreasing $\Delta/\gamma$.

To determine whether the real part of steady-state Fano coherence can reach a maximum as a function of $\Delta/\gamma$ and $\bar{n}$, we calculate the first derivatives of Eq.~(\ref{CanEqn_Imag}) with respect to these parameters
\begin{align}
    \frac{\partial }{\partial \bar{n}}\rho^{R}_{ab} &= \frac{(3\bar{n}^{2}+2\bar{n}+1)(\Delta/\gamma)^{2}+(\bar{n}^{2}+2\bar{n}+1)}{[(3\bar{n}+1)(\Delta/\gamma)^{2}+(4\bar{n}^{2}+5\bar{n}+1)]^{2}} \nonumber \\
    \frac{\partial }{\partial (\Delta/\gamma)}\rho^{R}_{ab} &= -\frac{2\bar{n}(\bar{n}+1)(\Delta/\gamma)}{[(3\bar{n}+1)(\Delta/\gamma)^{2}+(4\bar{n}^{2}+5\bar{n}+1)]^{2}}
\end{align} \par 
The derivatives do not vanish for $\bar{n}>0$ and $\Delta/\gamma>0$, and therefore  steady-state Fano coherences do not display maxima or minima. This behavior is illustrated in Fig.~\ref{fig:3}(c), which plots the real part of steady-state Fano coherence as a function of $\bar{n}$ and $\Delta/\gamma$. 
The magnitude of the steady-state Fano coherence is therefore maximal at a boundary of the two-dimensional region depicted in Fig.~\ref{fig:3}(c).  More specifically, the maximum is achieved in the small excited-state splitting regime $\Delta/\gamma\ll 1$ regardless of the value of $\bar{n}$. As the pumping rate increases, the region of maximum coherence is shifted to higher values of $\Delta/\gamma$. 

%To separate the effects of incoherent pumping from  enhancement of the steady-state coherences at strong pumping 
As an alternative coherence measure, consider the  ratio \cite{Tscherbul:14,Dodin:16}
\begin{equation}\label{CratioDef}
\mathcal{C} = \frac{\rho_{ab}^R}{\rho_{aa}},
\end{equation} 
which quantifies the relative magnitude of  coherences vs. excited-state populations. A substantial value of $\mathcal{C}\simeq 1$ indicates the possibility of a significant coherent contribution to excited-state  dynamics, affecting  the values of observables such as fluorescence emission intensities (see Sec. IV below).  From Eq. (\ref{SS_Dyn}) we obtain
\begin{equation}\label{Cratio1}
\mathcal{C} =  \frac{1}{1+\frac{\Delta^2}{\gamma(r+\gamma)} }
\end{equation} 
which is  less than unity as expected from the Cauchy-Swartz inequality, $|\rho_{ab}|^2\leq \rho_{aa}\rho_{bb}$. The ratio (\ref{Cratio1})  tends to unity  for $\frac{\Delta^2}{\gamma(r+\gamma)}\ll 1$, {\it i.e.}, in the regime {\it of small excited-state splitting  $\Delta$ compared to  $\sqrt{\gamma(r+\gamma)}$.} In the weak-pumping limit $(r\ll \gamma)$, this condition simplifies to  ${\Delta} \ll \gamma$. In the  strong-pumping limit ($r\gg  \gamma$),  steady-state Fano coherences are maximized relative to excited-state populations when ${\Delta} \ll \sqrt{r\gamma}$, which is a less restrictive condition than in the weak-pumping limit.

%We conclude that steady-state Fano coherences are maximized under the conditions of small excited-state splitting ($\Delta/\gamma \ll 1$) and strong  incoherent driving $(n\gg 1)$.  Interestingly, these are the same conditions as explored in previous theoretical studies \cite{Scully:11,Dorfman:13} of  noise-induced coherences in quantum heat engine models.

\subsection{Physical origin of steady-state Fano coherence: The two-bath nonequilibrium transport model}

Previous theoretical work has shown that Fano coherences induced in the V-system irradiated by isotropic incoherent radiation are transient and eventually decay back to zero \cite{Agarwal:01, Tscherbul:15, Dodin:16}. In contrast, when the V-system is driven by polarized incoherent radiation, Fano coherences remain non-zero in the steady state, suggesting the presence of a dynamic equilibrium  \cite{Agarwal:01,Dodin:18}.
 In our previous work we suggested an analogy between the latter situation and one of the V-system interacting with two independent thermal baths \cite{Dodin:18}, resulting in non-equilibrium heat transport across the V-system and steady-state coherences associated with it \cite{Segal:00,Manzano:13,Benenti:17}.  Here, we make this analogy more precise by showing that the BR equations of motion for the  V-system  driven by polarized incoherent radiation are equivalent to those describing quantum transport in a system consisting of two coupled two-level systems (qubits). We identify the imaginary part of steady-state Fano coherence with the non-equilibrium flux  across the dimer system. This result unravels a direct connection between the seemingly unrelated phenomena of incoherent driving of multilevel quantum systems and non-equilibrium transport.

\begin{figure}[t!]
\captionsetup{singlelinecheck = false, format= hang, justification=raggedright, font=footnotesize, labelsep=space}
\centering
\includegraphics[width=0.9\textwidth]{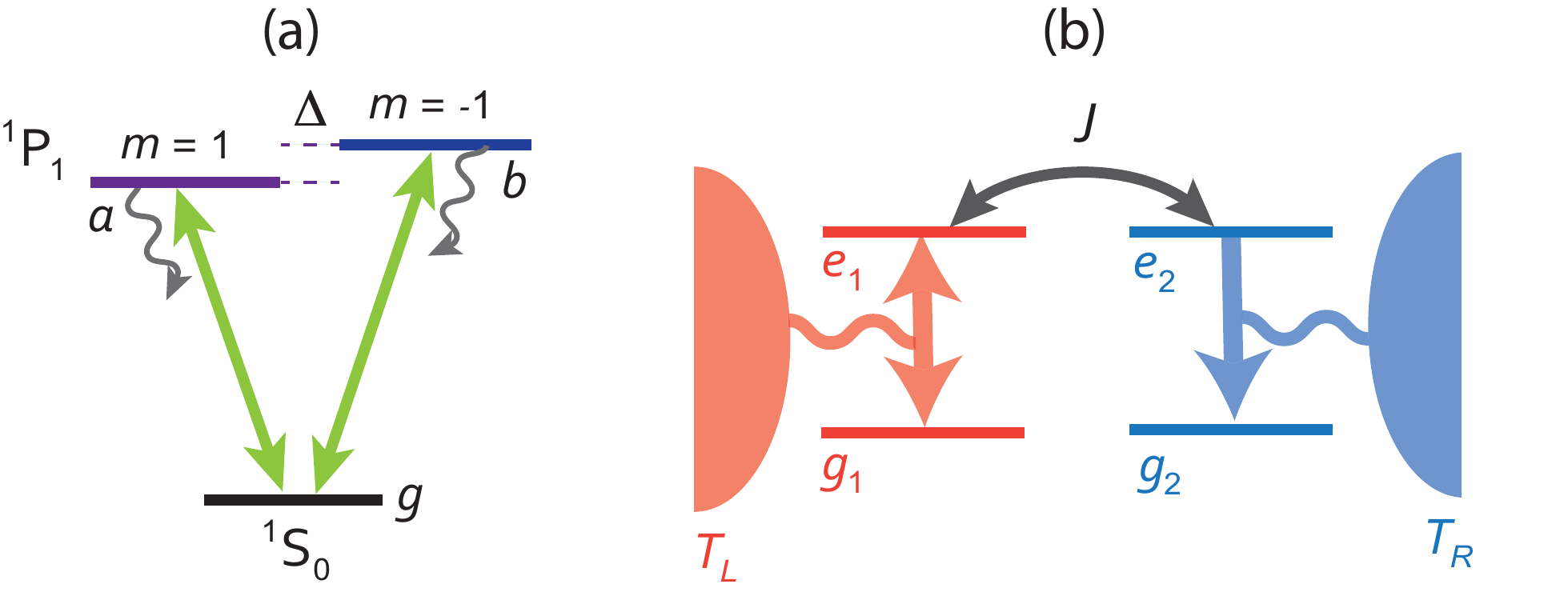}
\caption{(a) Energy level diagram of the V-system composed of a single ground state $g$ and two nearly degenerate excited levels $a$ and $b$. In our proposed experimental realization of the V-system, the ground state corresponds to the  $^1$S$_0$ level of atomic Ca and  the excited states corresponds to the  $^1$P$_1$ levels of  atomic Ca with the total electron orbital angular momentum $J=1$ and its projection $m_J=\pm 1$ on the magnetic field axis. The  excited-state energy splitting  $\Delta$ is continuously tunable with an external magnetic field. The rates of spontaneous decay $\gamma_i$ ($i = a,b$), and those of incoherent pumping  $r_i=\gamma_i \bar{n}$.  (b) An alternative representation of the V-system driven by polarized incoherent light as a pair of coherently coupled qubits interacting with two independent thermal baths at different temperatures.}
\label{fig:1}
\end{figure}

\begin{figure}[t!]
\captionsetup{singlelinecheck = false, format= hang, justification=raggedright, font=footnotesize, labelsep=space}
\centering
\includegraphics[width=0.95\textwidth]{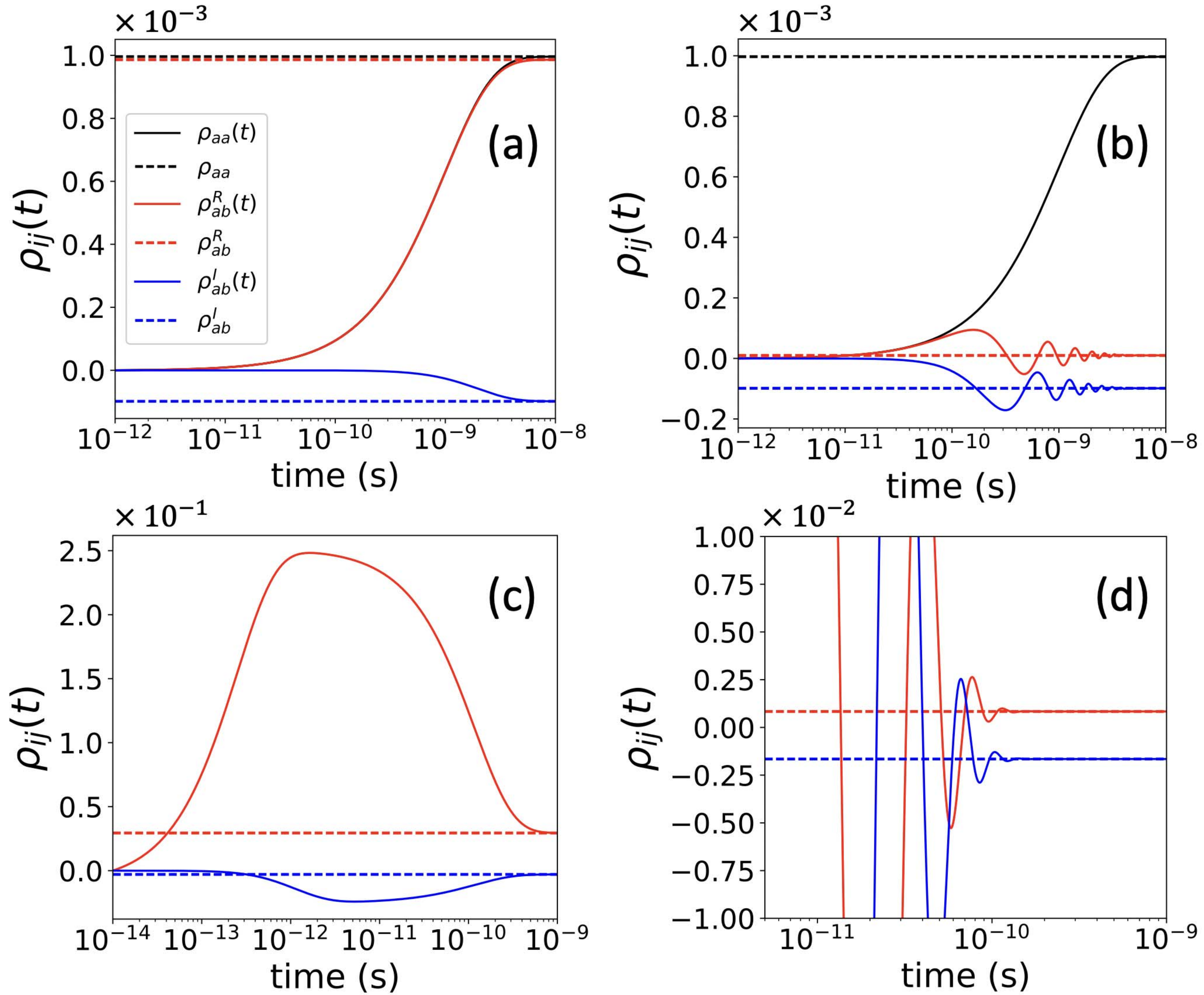}
\caption{Time evolution of  excited-state populations and Fano coherences in the V-system weakly (a)-(b) and strongly  (c)-(d) driven by $x$-polarized incoherent light.  (a) Overdamped regime with $\bar{n}=10^{-3}$ and $\Delta/\gamma= 10^{-1}$ (b) Underdamped regime with $\bar{n}=10^{-3}$ and $\Delta/\gamma= 10$;  (c)  $\bar{n}=10^{3}$ and $\Delta/\gamma= 10^{2}$, and (d) $\bar{n}=10^{2}$ and $\Delta/\gamma=2 \times 10^{2}$. The steady-state values of Fano coherences are shown by horizontal dashed lines.}
%All the initial population in the ground state ($i.e. \enspace \textbf{x}(0) = [\rho_{aa}(0), \rho^{R}_{ab}(0), \rho^{I}_{ab}(0)]^{T} = [0, 0, 0]^{T}$).}
\label{fig:2}
\end{figure}

\begin{figure}[t!]
\captionsetup{singlelinecheck = false, format= hang, justification=raggedright, font=footnotesize, labelsep=space}
\centering
\includegraphics[width=0.85\textwidth, trim = 0 270 0 260 ]{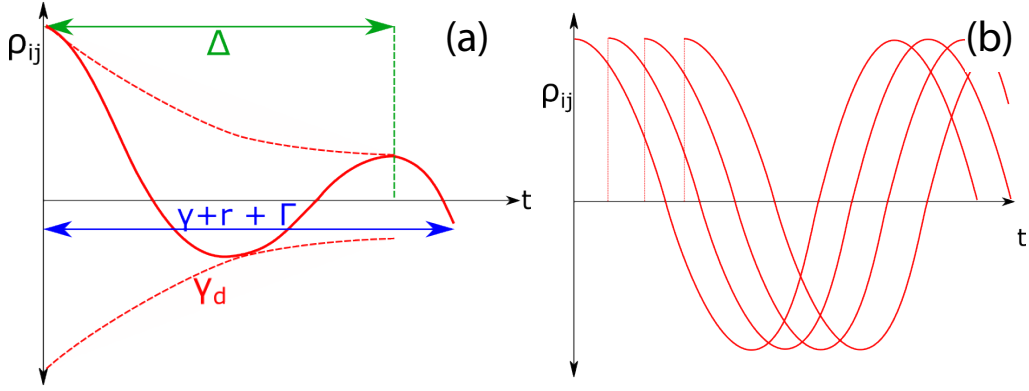}
\caption{Schematic depiction of the unitary time evolution and decay of a single incoherent excitation (a) and of an ensemble of incoherent excitations (b).}
\label{fig:timescales_dephasing}
\end{figure}

\begin{figure}[t!]
\captionsetup{singlelinecheck = false, format= hang, justification=raggedright, font=footnotesize, labelsep=space}
\centering
\includegraphics[width=1.05\textwidth]{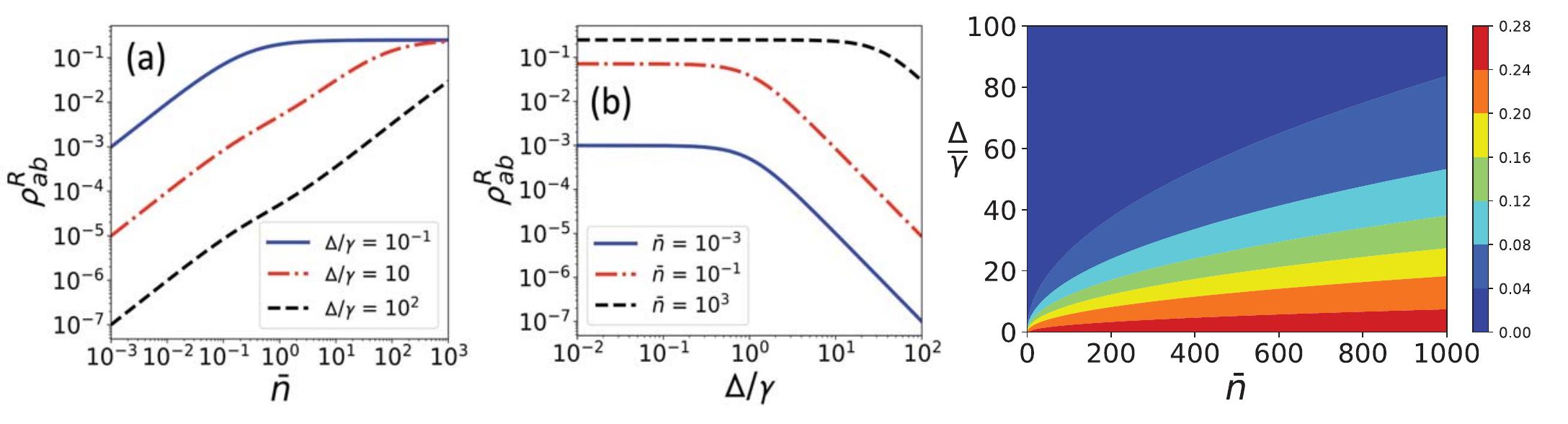}
\caption{(a) Steady-state Fano coherence plotted as a function of the average number of thermal photons $\bar{n}$ at fixed values of the excited-state splitting $\Delta/\gamma$.  (b) Same as in panel (a) plotted vs $\Delta/\gamma$ at fixed values of  $\bar{n}$ corresponding to the weak, intermediate, and strong pumping regimes.  (c) Two-dimensional contour plot of steady-state Fano coherence as a function of $\Delta/\gamma$ and $\bar{n}$.}
\label{fig:3}
\end{figure}

Consider a system of two qubits coherently coupled by an exchange or dipolar interaction quantified by the hopping  parameter $t=2\Delta$, as shown in Fig.~\ref{fig:1}(b).  In the weak-pumping limit only the ground $|g\rangle=|g_1g_2\rangle$ and singly excited eigenstates
\begin{align}\notag
|a\rangle&=\frac{1}{\sqrt{2}}(|e_1g_2\rangle+|e_2g_1\rangle), \\
|b\rangle&=\frac{1}{\sqrt{2}}(|e_1g_2\rangle-|e_2g_1\rangle) \label{TLSeigenstates}
\end{align}
of the two-qubit system are appreciably populated, leading to an effective V-system  description \cite{Tscherbul:18,Li:15} illustrated in  Fig.~\ref{fig:1}(a). Here $|g_i\rangle$ and $|e_i\rangle$  stand for the ground and excited computational basis states of the $i$-th  qubit. In addition, each qubit is coupled to a thermal bath maintained at different temperatures $T_L$ and $T_R$ as shown in Fig.~\ref{fig:1}(b). The perturbative BR master equations describing non-equilibrium transport of energy through the two-qubit system are, in the eigenstate basis  of the coupled two-qubit system (\ref{TLSeigenstates})  \cite{Li:15}
\begin{align}
    \Dot{\rho}_{aa} &= 2\Gamma^{+}_{aa}(\varepsilon_{a}) \rho_{gg} - 2\Gamma^{-}_{aa}(\varepsilon_{a}) \rho_{aa} - \Gamma^{-}_{ab}(\varepsilon_{b}) \rho_{ab} - \Gamma^{-}_{ba}(\varepsilon_{b}) \rho_{ba} \nonumber \\
    \Dot{\rho}_{bb} &= 2\Gamma^{+}_{bb}(\varepsilon_{b}) \rho_{gg} - 2\Gamma^{-}_{bb}(\varepsilon_{b}) \rho_{bb} - \Gamma^{-}_{ab}(\varepsilon_{a}) \rho_{ab} - \Gamma^{-}_{ba}(\varepsilon_{a}) \rho_{ba} \nonumber \\
    \Dot{\rho}_{ab} &= \big[\Gamma^{+}_{ba}(\varepsilon_{a}) \rho_{gg} - \Gamma^{-}_{ba}(\varepsilon_{a}) \rho_{aa}\big] +  \big[\Gamma^{+}_{ba}(\varepsilon_{b}) \rho_{gg} - \Gamma^{-}_{ba}(\varepsilon_{b}) \rho_{bb}\big] + i\Delta \rho_{ab} - \big[\Gamma^{-}_{aa}(\varepsilon_{a}) + \Gamma^{-}_{bb}(\varepsilon_{b})\big] \rho_{ab} \label{QME_Sun}
\end{align}
Significantly,  these equations describe the setup, where qubit 1 is  coupled {\it only} to the left bath and qubit 2 {\it only} to the right bath, as can be shown by recasting the system-bath coupling given by Eq. (A.1) of Ref. \cite{Li:15} in the site basis via Eq. (\ref{TLSeigenstates}).
In Eq. (\ref{QME_Sun}), $\varepsilon_{i} = E_{i} - E_{g}$ is the energy difference between the excited state $i = \ket{a},\enspace \ket{b}$ and the ground state $\ket{g}$, $\Delta = E_{a} - E_{b}$ is the excited state splitting and the dissipation rates $\Gamma^{\pm}_{ij}(\omega)$ are  given by
\begin{align}
    \Gamma^{-}_{ij}(\omega) &= \frac{1}{2} \gamma^{(L)}_{ij}(\omega) \big[\bar{n}_{L}(\omega)+1 \big] + \frac{1}{2} \gamma^{(R)}_{ij}(\omega) \big[\bar{n}_{R}(\omega)+1 \big] \nonumber \\
    \Gamma^{+}_{ij}(\omega) &= \frac{1}{2} \gamma^{(L)}_{ij}(\omega) \bar{n}_{L}(\omega) + \frac{1}{2} \gamma^{(R)}_{ij}(\omega) \bar{n}_{R} \label{QME_Sun2},
\end{align}
$\bar{n}_{\alpha} = (e^{\hbar\omega/T_{\alpha}}-1)^{-1}$  is the average number of photons in the  left and right baths $\alpha=L,R$, and  $\gamma^{\alpha}_{ij}(\omega)$ are the coupling spectra of the $\alpha$-th bath 
\begin{equation}
    \gamma^{\alpha}_{ij}(\omega) = 2\pi \sum_{k_{\alpha}} g^{*}_{i,k_{\alpha}} g_{j,k_{\alpha}} \delta(\omega-\omega_{k_{\alpha}})= [\gamma^{\alpha}_{ji}(\omega)]^{*} 
\end{equation}
where $g_{i, k_{\alpha}} = \sqrt{\frac{\hslash \omega_{k_{\alpha}}}{2 \epsilon_{0} V}} \big(\bm{\mu}_{ij} \cdot \bm{\varepsilon}_{k_{\alpha} \lambda} \big)$ is the light-matter coupling coefficient. The cross specta $\gamma^{\alpha}_{ij}$ ($i \neq j$)  quantify the interference between the transitions $\ket{a} \rightarrow \ket{g}$ and $\ket{b} \rightarrow \ket{g}$   \cite{Li:15}
\begin{equation}\label{weight_factors}
    |\gamma^{\alpha}_{ij}(\omega)|^{2} = f_{\alpha}(\omega)  \gamma^{\alpha}_{ii}(\omega) \gamma^{\alpha}_{jj}(\omega), 
\end{equation}
where the weight factors $f_{\alpha}$ (assumed henceforth to be frequency independent) are determined by  the form of the system-bath coupling \cite{Li:15}. In the case of incoherent light-matter coupling of interest here, the weight factors are directly related to the transition dipole alignment parameters $p_{ij}$ \cite{Tscherbul:15}. 

We suggest that the V-system irradiated by polarized incoherent radiation can be thought of as a system of two qubits interacting with each other and with two independent thermal baths, as shown in Fig.~\ref{fig:1}(b). A hot bath with $T_L=5800$~K is responsible for incoherent pumping of the qubit coupled to it, which allows us to set $\bar{n}_{L} = \bar{n}$  in Eq. (\ref{QME_Sun}).   A cold bath with $T_R=0$ accounts for spontaneous emission due to the coupling of the second qubit to the vacuum modes of the electromagnetic field, and hence we can set $\bar{n}_{R}=0$ in Eq. (\ref{QME_Sun}).
With these simplifications, Eqs. (\ref{QME_Sun}) and (\ref{QME_Sun2}) reduce to
\small
\begin{align}
    \Dot{\rho}_{aa} &= r_{aa}^{L} \rho_{gg} - (r_{aa}^L+{\gamma}_{aa}^L + {\gamma}_{aa}^R) \rho_{aa} - \frac{1}{2} \sqrt{f_{L}}\big(\sqrt{r_{aa}^Lr_{bb}^L}+\sqrt{{\gamma}_{aa}^L{\gamma}_{bb}^L}\big) (\rho_{ab}+\rho_{ba}) - \frac{1}{2} \sqrt{f_{R}} \sqrt{{\gamma}^R_{aa}{\gamma}^R_{bb}} (\rho_{ab}+\rho_{ba}) \nonumber \\
        \Dot{\rho}_{bb} &= r_{bb}^{L} \rho_{gg} - (r_{bb}^L+{\gamma}_{bb}^L + {\gamma}_{bb}^R) \rho_{bb} - \frac{1}{2} \sqrt{f_{L}}\big(\sqrt{r_{aa}^Lr_{bb}^L}+\sqrt{{\gamma}_{aa}^L{\gamma}_{bb}^L}\big) (\rho_{ab}+\rho_{ba}) - \frac{1}{2} \sqrt{f_{R}} \sqrt{{\gamma}^R_{aa}{\gamma}^R_{bb}} (\rho_{ab}+\rho_{ba}) \nonumber \\
 %   \Dot{\rho}_{bb} &= r_{b} \rho_{gg} - (r_{b}+\tilde{\gamma}_{b}+\tilde{\gamma}'_{b}) \rho_{bb} - \frac{1}{2} \sqrt{f_{L}(\omega)}\big(\sqrt{r_{a}r_{b}}+\sqrt{\tilde{\gamma}_{a}\tilde{\gamma}_{b}}\big) (\rho_{ab}+\rho_{ba}) - \frac{1}{2} \sqrt{f_{R}(\omega)} \sqrt{\tilde{\gamma}'_{a}\tilde{\gamma}'_{b}} (\rho_{ab}+\rho_{ba}) \\
\begin{split}
    \Dot{\rho}_{ab} &= \sqrt{f_{L}} \sqrt{r_{aa}^Lr_{bb}^L} \rho_{gg} - \frac{1}{2} \sqrt{f_{L}}\big(\sqrt{r_{aa}^Lr_{bb}^L}+\sqrt{{\gamma}^L_{aa}{\gamma}_{bb}^L}\big) (\rho_{aa}+\rho_{bb}) - \frac{1}{2} \sqrt{f_{R}} \sqrt{{\gamma}^R_{aa}{\gamma}^R_{bb}} (\rho_{aa}+\rho_{bb}) - i\Delta \rho_{ab} \\ &\qquad - \frac{1}{2} (r_{aa}^L+r_{bb}^L+{\gamma}^L_{aa}+{\gamma}_{bb}^L+{\gamma}_{aa}^R+{\gamma}^R_{bb}) \rho_{ab},
\end{split} \label{BR_Transport_Eqn}
\end{align}\normalsize
where we use the same incoherent pumping and spontaneous decay parameters $\gamma^{\alpha}_{ij}$ and $r^{\alpha}_{ij}$ ($ i = a, b$;  $\alpha = L, R$) as the transport model equations (\ref{QME_Sun}) and (\ref{QME_Sun2}) \cite{Li:15}. 
%Here, $\tilde{\gamma}_{a} = \gamma^{L}_{aa}$, $\tilde{\gamma}_{b} = \gamma^{L}_{bb}$, $\tilde{\gamma}'_{a} = \gamma^{R}_{aa}$, $\tilde{\gamma}'_{b} = \gamma^{R}_{bb}$ are the spontaneous decay rates. 
The  incoherent pumping rates due to the left (hot) bath are given by $r_{ii}^L = \bar{n}_L {\gamma}^L_{ii}$, where ${\gamma}^L_{ii}$ are the corresponding spontaneous decay rates. The right (cold) bath can only induce spontaneous decay, whose rates are given by  ${\gamma}^R_{ii}$. %of the light-matter coupling coefficient $g_{i,k_{\alpha}}$. 

In this paper, we consider the V-system with orthogonal transition dipole moments (such as the Ca atom in Sec.~IV  irradiated by polarized incoherent radiation.  Because of the orthogonality condition, there is no interference in spontaneous emission  \cite{Dodin:18} and we can thus set $f_{\alpha}= 0$ for spontaneous emission terms in Eq. (\ref{BR_Transport_Eqn}).  In contrast, we have $f_{\alpha} = 1$  for polarized incoherent driving \cite{Dodin:18}. As a result, Eq. ~(\ref{BR_Transport_Eqn}) can be recast in the form 
\begin{align}
    \Dot{\rho}_{aa} &= r_{aa}^{L} \rho_{gg} - (r_{aa}^L+{\gamma}_{aa}^L + {\gamma}_{aa}^R) \rho_{aa} - \frac{1}{2} \sqrt{r_{aa}^Lr_{bb}^L}(\rho_{ab}+\rho_{ba})  \nonumber \\
        \Dot{\rho}_{bb} &= r_{bb}^{L} \rho_{gg} - (r_{bb}^L+{\gamma}_{bb}^L + {\gamma}_{bb}^R) \rho_{bb} - \frac{1}{2}   \sqrt{r_{aa}^Lr_{bb}^L} (\rho_{ab}+\rho_{ba}) \nonumber \\
    \Dot{\rho}_{ab} &= \sqrt{r_{aa}^Lr_{bb}^L} \rho_{gg} - \frac{1}{2}  \sqrt{r_{aa}^Lr_{bb}^L} (\rho_{aa}+\rho_{bb}) - i\Delta \rho_{ab}  - \frac{1}{2} (r_{aa}^L+r_{bb}^L+{\gamma}^L_{aa}+{\gamma}_{bb}^L+{\gamma}_{aa}^R+{\gamma}^R_{bb}) \rho_{ab}
\label{BR_Transport_Eqn_set_factors}
\end{align}\normalsize

These equations are identical to the BR quantum optical master equation (\ref{QME1}), with $\gamma_{i} = {\gamma}_{ii}^L + {\gamma}^R_{ii}$ and $r_i=r_{ii}^L=\bar{n}_L\gamma^L_{ii}$,  thus establishing equivalence between the V-system driven by polarized incoherent light and a qubit dimer coupled to two independent baths.  
Significantly, the transport model gives the ratio of incoherent driving and spontaneous decay rates ${r_i}/\gamma_i=r_{ii}^L/(\gamma_{ii}^L + \gamma_{ii}^R)$ at odds with the equilibrium fluctuation-dissipation theorem, based on which one would expect ${r_i}/\gamma_i = r_{ii}^L/\gamma_{ii}^L=\bar{n}_L$. As discussed above and in Refs.~\cite{Agarwal:01,Dodin:19}, it is this  imbalance between incoherent driving and spontaneous decay   that leads to   the formation of coherent NESS.

%\begin{align}
%    \Dot{\rho}_{ii} &= r_{i} \rho_{gg} - (r_{i}+\tilde{\gamma}_{i} + \tilde{\gamma}'_{i}) \rho_{aa} - \frac{1}{2} \sqrt{r_{a}r_{b}} (\rho_{ab}+\rho_{ba}) \nonumber \\
%    \Dot{\rho}_{ab} &= \sqrt{r_{a}r_{b}} \rho_{gg} - \frac{1}{2} \sqrt{r_{a}r_{b}} (\rho_{aa}+\rho_{bb}) - i\Delta \rho_{ab} - \frac{1}{2} (r_{a}+r_{b}+\tilde{\gamma}_{a}+\tilde{\gamma}_{b}+\tilde{\gamma}'_{a}+\tilde{\gamma}'_{b}) \rho_{ab} \label{BR_Transport_Eqn2}
%\end{align}

%. The $p$ factor for the isotropic spontaneous emission process is quantified by the transition dipole alignment factor $p$ \cite{Tscherbul:14, Tscherbul:15, Dodin:16}. For the Calcium atom V-system, we have $p = 0$ \cite{Dodin:18}. However, the light-matter coupling term for the stimulated emission process due to the linearly polarized incoherent radiation is evaluated as $\sqrt{r_{a}r_{b}}$. With these observations, 
%$f_{\alpha} = 1$ for stimulated emission process and 

To characterize non-equilibrium transport in our model two-qubit system, we calculate the energy flux from the ``hot'' qubit 1 to the ``cold'' qubit 2 \cite{Manzano:13,Li:15}
\begin{equation}
J_{1-2} = -i \langle [\hat{\sigma}^{z}_{1}, \hat{H}_{s}] \rangle =
         -i 2g \text{Tr} \big[ \rho \big(\ket{a} \bra{b} - \ket{b} \bra{a}\big)\big] = 4g \textrm{Im}\{\langle a|\rho | b \rangle \},\label{J1_to_2}
\end{equation}
where $\hat{H}_{s} = \frac{1}{2} \omega_{1}\hat{\sigma}^{z}_{1} + \frac{1}{2} \omega_{2}\hat{\sigma}^{z}_{2} + g (\hat{\sigma}^{+}_{1}\hat{\sigma}^{-}_{2} + \hat{\sigma}^{-}_{1}\hat{\sigma}^{+}_{2})$ is the  Hamiltonian of the  two-qubit system expressed via the jump operators $\hat{\sigma}_i^\pm$ which are defined as $\hat{\sigma}_i^{-} = \ket{g_{i}} \bra{e_{i}}$, $\hat{\sigma}_i^{+} = \ket{e_{i}} \bra{g_{i}}$, $\hat{\sigma}_i^{z} = \ket{e_{i}} \bra{e_{i}} - \ket{g_{i}} \bra{g_{i}}$. Here, $\ket{e_{i}}$, $\ket{g_{i}}; \enspace i = 1,2$ are the excited and ground states of the  coupled qubits 1 and 2 in the site basis, $g$ is the qubit-bath coupling, and the trace is taken over the bath degrees of freedom.   

Equation (\ref{J1_to_2}) establishes that the energy flux  from  qubit 1 to qubit 2 is proportional to the  imaginary part of Fano coherence in the energy eigenstate basis (\ref{TLSeigenstates}) of the dimer. This clarifies the physical significance of steady-state Fano coherences as being responsible for radiative energy transfer from sunlight-driven qubit 1 to qubit 2 that is  coupled to the vacuum modes of the electromagnetic field. In addition, the real part of coherence in the energy eigenstate basis is proportional to the difference of excited state populations in the site basis \cite{Tscherbul:18}. %Mathematically, it is evaluated as $\rho^{R}_{ab} = \frac{1}{2} (\rho_{e_{1}e_{1}} - \rho_{e_{2}e_{2}})$ where, $\rho_{e_{i}e_{i}} = \bra{e_{i}} \rho \ket{e_{i}}; \enspace i = 1, 2$. 

\subsection{Effect of excited-state relaxation and decoherence}

Realistic quantum systems are always subject to environmental perturbations causing relaxation and decoherence. In this section, we explore the robustness of steady-state Fano coherences against these detrimental effects. To this end, we follow previous theoretical work \cite{Rebentrost:09,Chin:09,Tscherbul:18} and introduce phenomenological relaxation and decoherence channels represented by the terms $\Gamma \rho_{aa}$ and $\gamma_d\rho_{ab}$ in Eq. (\ref{QME2}). 
% for $\rho_{ab}$.
%  I assume that environmental relaxation is negligibly slow compared to the rates of the radiative processes ($\Gamma = 0$). Setting $\gamma_{d} \neq 0$ in the BR Eqs.~(\ref{QME}) and 
Following the analysis presented above, we obtain steady state solutions as a function of relaxation and decoherence rates $\Gamma$ and $\gamma_d$:
\begin{align} \label{Dyn_Dec_Rel}
    \rho_{aa} &= \frac{r [\Delta^{2}+(\gamma + \gamma_{d})(r+\gamma+\gamma_{d})]} {(3r+\gamma+\Gamma)[\Delta^{2}+(r+\gamma+\gamma_{d})^{2}]-3r^{2}(r+\gamma+\gamma_{d})} \nonumber \\
    \rho^{R}_{ab} &= \frac{r (\gamma+\Gamma) (r+\gamma+\gamma_{d})}{(3r+\gamma+\Gamma)[\Delta^{2}+(r+\gamma+\gamma_{d})^{2}]-3r^{2}(r+\gamma+\gamma_{d})} \nonumber \\
    \rho^{I}_{ab} &= -\frac{\Delta}{(r+\gamma+\gamma_{d})} \rho^{R}_{ab}
\end{align} \par 
These equations suggest that both the real and imaginary steady state coherences can survive in the presence of relaxation and decoherence. Increasing the relaxation rate does not significantly affect the steady-state coherence in the typical limit  $\Gamma \gg r+\gamma$, where the dependence $\rho^{R}_{ab}(\Gamma) \simeq \frac{\gamma+\Gamma}{3r+\gamma+\Gamma}$ is weak (Ref.~\cite{KoyuThesis:19} considers the effect of relaxation is more detail). We therefore focus on the case of pure decoherence (or dephasing) assuming $\Gamma=0$.  
%The effect of relaxation is always to diminish the value of steady-state populations [since the relaxation rate $\Gamma$ only enters through the prefactor $\frac{1}{3r+\gamma+\Gamma}$ in Eq. (\ref{Dyn_Dec_Rel})], the role of decoherence is more subtle. To isolate this role, we will henceforth focus on the case of pure decoherence ($\Gamma=0$).  
%The effects of nonradiative relaxation  on the steady-state Fano coherence in polarized incoherent excitation %, where steady-state Fano coherences were shown to vanish for the symmetric dimer undergoing symmetric relaxation ($\Gamma_1=\Gamma_2$).

% \textcolor{RubineRed}{It is instructive to compare Eqs. (\ref{SS_Dyn}) with Eqs. (15) of Ref. \cite{Dodin:19}, whoch predict no steady-state coherences in the energy eigenstate basis for the symmetric relaxation ($\Gamma_1=\Gamma_2$). )

\begin{figure}[t!]
\captionsetup{singlelinecheck = false, format= hang, justification=raggedright, font=footnotesize, labelsep=space}
\centering
\includegraphics[width=1.05\textwidth]{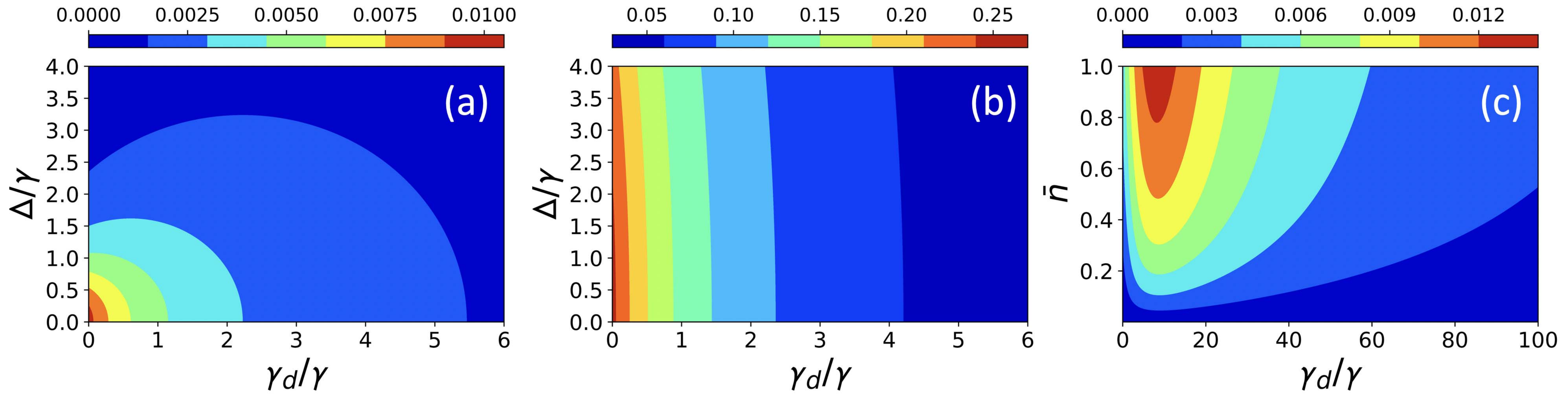}
\caption{Contour plots of steady-state Fano coherence as a function of the excited-state splitting $\Delta/\gamma$ and of the decoherence rate $\gamma_d/\gamma$ under weak (a) and strong (b) $x$-polarized incoherent pumping. The values of the average photon number used are $\bar{n}=0.01$ (a)  and $\bar{n}=100$ (c).  Panel (c) shows the steady-state Fano coherence as a function the decoherence rate and $\bar{n}$ at a fixed $\Delta/\gamma=10$. }
\label{fig:4}
\end{figure}

In Fig.~\ref{fig:4} we plot the dependence of the steady-state Fano coherence on the excited-state splitting and the reduced decoherence rate $\gamma_d/\gamma$  (\ref{Dyn_Dec_Rel}).  We observe that, regardless of the pumping intensity, the steady-state coherences are maximized for small excited-state splittings $\Delta/\gamma \ll 1$. In the weak pumping limit illustrated in Fig.~\ref{fig:4}(a) the coherences also decrease rapidly with the decoherence rate, falling below one part in $10^{3}$ for $\gamma_d/\gamma > 6$. Similar trends are observed in the case of intermediate pumping ($\bar{n}=1$), even though the absolute magnitude of the steady-state coherence in this regime is larger by two orders of magnitude.  The coherences in the strong pumping limit ($\bar{n}\gg 1$) are large and while they they do not fall off as strongly with $\Delta/\gamma$ they do get significantly suppressed  by decoherence. The extent of this suppression appears insensitive to the pumping rate.

Remarkably, as shown in Figs.~\ref{fig:4}(a)-(c),  the presence  of decoherence can {\it enhance} the magnitude of steady-state Fano coherences.  For instance, at $\Delta/\gamma_d=2$ the value of $\rho_{ab}^{R}$ passes through a maximum near $\gamma_d=2$ and then decreases again at higher $\gamma_d$.  This enhancement  is particularly pronounced for large $\Delta/\gamma$ in the weak-pumping limit, where we observe up to 5-fold coherence enhancements  \cite{KoyuThesis:19}.   To further explore this counterintuitive effect, we rewrite Eq.~(\ref{Dyn_Dec_Rel}) in terms of the dimensionless parameters  $\Tilde{\gamma}_{d} = \gamma_{d}/\gamma$ and $\Tilde{\Delta} = \Delta/\gamma$
\begin{equation} \label{SS_Real}
    \rho^{R}_{ab} = \frac{\bar{n} (\bar{n}+1+\Tilde{\gamma}_{d})}{(3\bar{n}+1)[\Tilde{\Delta}^{2}+(\bar{n}+1+\Tilde{\gamma}_{d})^{2}]-3\bar{n}^{2}(\bar{n}+1+\Tilde{\gamma}_{d})}
\end{equation}
The roots of the equation $ \frac{d}{d\Tilde{\gamma}_{d}}\rho_{ab}^{R}=0$  are given by  $\Tilde{\gamma}_{d} = \pm \tilde{\Delta}  -\bar{n} - 1$   \cite{KoyuThesis:19}. Picking the physical root corresponding to $\tilde{\Delta}>0$, we obtain the optimal decoherence rate
\begin{equation}\label{gammadRoot}
  \gamma_{d} =  {\Delta}  -{r} - \gamma
  \end{equation}
  that maximizes the value of the steady-state Fano coherence.
In the weak pumping regime, Eq.~(\ref{gammadRoot})  simplifies to  ${\gamma}_{d} \approx {\Delta}$ which defines the diagonal  ``line of optimal decoherence'' in Figs.~\ref{fig:4}(a) and (b), along which the steady-state Fano coherence takes a maximum value. Importantly, the maximum of  the steady-state Fano coherence as a function of $\gamma_d$ only appears when the optimal decoherence rate in Eq. (\ref{gammadRoot}) is positive (${\Delta}\ge r+\gamma$), which  corresponds to the underdamped regime of V-system dynamics \cite{Koyu:18,Tscherbul:14,Dodin:16,Dodin:18}.
This is illustrated in Fig.~\ref{fig:4}(c), which shows that the maximum of $\rho_{ab}^R$ is located along the vertical line  of constant $\gamma_d/\gamma=10$.

%To separate the effects of incoherent pumping from  enhancement of the steady-state coherences at strong pumping 
The population-to-coherence ratio (\ref{CratioDef}) in the presence of dephasing can be obtained from Eq. (\ref{Dyn_Dec_Rel}) 
\begin{equation}
\mathcal{C} = \frac{1}{1+ \frac{\gamma_d}{\gamma} + \frac{\Delta^2}{\gamma(r+\gamma+\gamma_d)} }.
\end{equation} 
The ratio takes the maximum value $\mathcal{C}_\text{max}=\frac{1}{1+\gamma_d/\gamma + (r+\gamma+\gamma_d)/\gamma }$ when the sum in the denominator 
%($\frac{\gamma_d}{\gamma} + \frac{\Delta^2}{\gamma(r+\gamma+\gamma_d)}$) 
is a minimum, which occurs when $\gamma_d=\Delta-r-\gamma$. This is the same condition as required for the maximization of the real part of steady-state Fano coherence (\ref{gammadRoot}). 

 Under typical molecular excitation  conditions, nonradiative dephasing is fast compared with radiative processes ($\gamma_d\gg \gamma, r$)  and we obtain $\mathcal{C} \simeq \frac{1}{1+\gamma_d/\gamma + \Delta^2/(\gamma\gamma_d) }\ll 1$. In the limit of small $\Delta\ll \sqrt{\gamma\gamma_d}$ the last term in the denominator can be neglected, giving 
  $\mathcal{C} \simeq \frac{1}{1+\gamma_d/\gamma }\simeq \gamma/\gamma_d \ll 1$.
   Thus, {\it  the ratio of steady-state Fano coherences to excited-state populations will be suppressed by the factor $\gamma_d/\gamma$ in the presence of environmental dephasing, as long as the system-environment coupling responsible for the dephasing does not generate any steady-state coherence} (as assumed here, see Sec. III).

\begin{figure}[t!]
\captionsetup{singlelinecheck = false, format= hang, justification=raggedright, font=footnotesize, labelsep=space}
\centering
\includegraphics[width=0.85\textwidth]{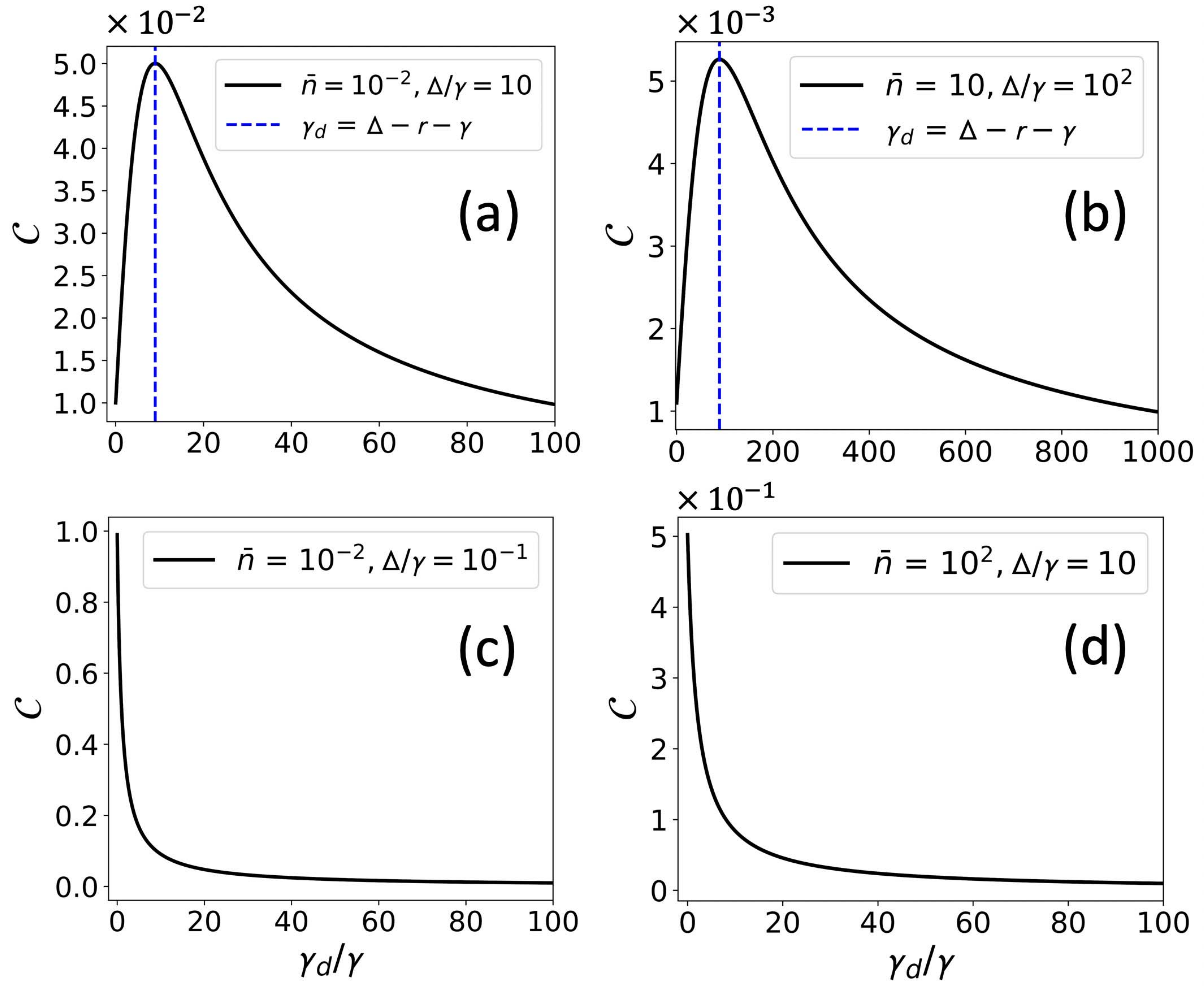}
\caption{Steady-state coherence-to-population  ratio $\mathcal{C}$ plotted as a function of dimensionless decoherence rate  $\gamma_d/\gamma$ in the weak-pumping limit [(a), (c)] and  in the strong-pumping limit [(b), (d)].  The values of  the $\Delta/\gamma$ are  indicated in each panel.  }
\label{fig:Cratio}
\end{figure}

Figure~\ref{fig:Cratio} shows the steady-state coherence-to-population ratio as a function of dephasing. In accordance with  the analytic result obtained above (\ref{gammadRoot}), we observe the emergence of a maximum at the optimal decoherence rate $\gamma_d=\Delta-r-\gamma$. The maximum is only present at sufficiently large excited-state splittings $\Delta$ such that $\Delta-r-\gamma > 0$ (underdamped regime of V-system dynamics  \cite{Tscherbul:14,Dodin:16a,Koyu:18}) as shown in Figs.~\ref{fig:Cratio}(a) and (b).   Decreasing the value of $\Delta$ brings the V-system into the overdamped regime \cite{Tscherbul:14,Dodin:16a,Koyu:18} characterized by $\Delta/\gamma<1$. As shown in Figs.~\ref{fig:Cratio}(c) and (d), this leads to the disappearance of the maximum and the  expected monotonic decrease of  the $\mathcal{C}$-ratio with increasing  decoherence rate.

While the ability of dephasing to enhance steady-state coherence in the  underdamped regime is surprising, the dephasing enhanced steady-state coherence can be understood by considering the individual contributing excitations.
In the underdamped regime, one mechanism of coherence loss arises due to the cancellation of excitations at different times that carry different phases due to their unitary evolution in the excited state manifold [see Fig~\ref{fig:timescales_dephasing}(b)]. By introducing an explicit dephasing channel, excitations generated at earlier times lose more coherences than those generated later. As  a result, these excitations do not destructively interfere  as much as they could in the absence of dephasing. The dephasing then weights the phase average towards a narrower window of excitation times, decreasing the ensemble-induced dephasing process. However, dephasing leads to coherence loss at the level of individual excitations.
Consequently, this trade-off leads to a competition between the two effects of increased dephasing, allowing for dephasing-enhanced steady state coherences in certain regimes when the decreased ensemble dephasing outweighs the increased excitation level dephasing.

We conclude that in the underdamped (or large-$\Delta$) regime of V-system dynamics, steady-state  Fano coherence are maximized when the decoherence rate is equal to the energy splitting between the excited states.  The physical origin of the maximization can be interpreted at the level of individual incoherent excitations, whose unitary evolution is accompanied by environmental decoherence. The competition of these mechanisms can result in an enhancement of steady-state coherence through minimization of destructive interference caused by coherent evolution. A similar mechanism is at play in environment-assisted quantum transport, whereby the addition of a moderate amount of noise positively affects the energy flux in  multichromophoric  reaction complexes \cite{Manzano:13,Wu:10,Rebentrost:09,Chin:09}.
 The decoherence-induced enhancement of Fano coherences could be used to facilitate their  experimental observation via measuring excited-state populations, as discussed in the following section.

% of the steady-state Fano coherences.
%Fig (\ref{fig: SS_rhoabR_vs_gammad}b) illustrates the maximum of the real part of coherences in the weak pumping regime with $\bar{n} = 10^{-2}$ and  $\Delta/\gamma = 10$. 

%Taking the first derivative with respect to $\Tilde{\gamma}_{d}$ and solving t

%This effect is counterintuitive because one would normally expect decoherence to destroy any coherence present in the system. 
%Particularly interesting is the large $\Delta/\gamma$ regime, we observe an unexpected increase of steady-state coherence as a function of $\gamma_{d}$ at low pumping intensities between the curve for $\gamma_{d} = 0$ and $\gamma_{d} = 10\gamma, \enspace 50\gamma$.
%ven in the presence of rapid decoherence with $\gamma_{d} = 50\gamma$. This is a remarkable result as we would normally expect the decoherence effect to destroy any residual coherences. Particularly interesting is the large $\Delta/\gamma$ regime, we observe an unexpected increase of steady-state coherence as a function of $\gamma_{d}$ at low pumping intensities between the curve for $\gamma_{d} = 0$ and $\gamma_{d} = 10\gamma, \enspace 50\gamma$. \par 

\section{Experimental detection of steady-state Fano coherence }

In our previous work \cite{Dodin:18}, we proposed a scheme for the experimental observation of  {\it transient} Fano coherences by detecting  the fluorescence signal emitted by incoherently pumped Calcium atoms. This scheme suffers from two  difficulties: First, the transient nature of noise-induced coherent dynamics makes it sensitive to the turn-on time of the radiation. In particular, Fano coherences are known to vanish in the limit where the turn-on time is much longer than that of spontaneous emission \cite{Dodin:16a}.
Second, because the total fluorescence intensity is independent of Fano coherence \cite{Dodin:18}, it required the detection of the fluorescence emitted into a specific range of solid angles.

In this section, we eliminate these difficulties by arguing that steady-state Fano coherences  can be observed by measuring the {\it total steady-state} fluorescence signal emitted by Calcium atoms irradiated by polarized vs. isotropic incoherent light.  Unlike the method proposed in Ref. \cite{Dodin:18}, the new detection scheme  requires neither rapid  turn-on of the radiation field nor spatially resolved detection of the fluorescence signal.

 Our proposed experimental setup is similar to that described in our previous work (see Fig.~\ref{fig:1}(a) and Fig. 1 of Ref.~\cite{Dodin:18}) and consists of a beam of Ca atoms initially prepared in the ground $^1$S$_0$ electronic state. The atoms are continuously  irradiated with spectrally broadened light polarized in the $x$-direction,  inducing the electric dipole transition $^1$S$_0\to {^1}\textrm{P}_{m_J}$ from the $J=0$ ground state to $J=1$ excited states with $m_J=\pm 1$. The excited states are split by an external magnetic field pointing in the $z$-direction, forming  a nearly perfect V-system with tunable excited-state splitting $\Delta$ \cite{Dodin:18}. Following initial turn-on of incoherent excitation, a fraction of Ca atoms in the beam will be excited to the $^1\textrm{P}_{m_J}$ states. The excited atoms will decay to the ground state emitting fluorescence photons.

 The average intensity of the radiation emitted by Ca atoms in the direction  $(\theta, \phi)$ at a distance $R$ away from the source is \cite{Dodin:18}
\begin{equation} \label{Int_Eq}
    \langle I(\mathbf{R},t) \rangle = \frac{n_{r}\omega^{4}_{0}}{32\pi^{2}
    \epsilon_{0} c^{3} R^{2}} \bigg[ \frac{1+\text{cos}^{2}\theta}{2} \big(\rho_{aa}(t') + \rho_{bb}(t')\big) + \text{sin}^{2}\theta \big(\text{cos}2\phi \rho^{R}_{ab}(t')-\text{sin}2\phi \rho^{R}_{ab}(t') \big) \bigg]
 \end{equation}
 where  $t' = t + R/c$ is the retarded time, $n_{r}$ is the refractive index of the medium, $\textit{c}$ is the speed of light in free space. Integrating Eq.~(\ref{Int_Eq}) over all spatial directions ($\theta, \phi$) we obtain the total emitted intensity $I = \frac{8\pi}{3} I^{0} (\rho_{aa} + \rho_{bb})$, where $I^{0} = n_{r}\omega^{4}_{0}/(32\pi^{2}\epsilon_{0} c^{3} R^{2})$. 
For a symmetric V-system $\rho_{aa}=\rho_{bb}$, this expression  further simplifies to
\begin{equation} \label{I_final2}
    I = \frac{16\pi}{3} I^{0} \rho_{aa}
\end{equation}
Importantly, unlike its spatially resolved counterpart (\ref{Int_Eq}) the integrated fluorescence intensity depends only on the excited state population and not on the Fano coherence. Therefore, in order to directly observe transient Fano coherences,  it is necessary to measure the intensity emitted into specific domains of space \cite{Dodin:18} in such a way that the partially integrated  fluorescence intensity is made to explicitly depend of Fano coherence terms $\rho_{ab}$.  

Here, we show that it is not necessary to implement spatially resolved fluorescence detection  \cite{Dodin:18} to observe steady-state Fano coherences in a V-system excited by polarized incoherent light. 
As shown in Sec. III above, the magnitude of the real part of steady-state Fano coherence is directly related to the {\it deviation of excited-state populations from their values in thermodynamic equilibrium} via Eq.~(\ref{CanEqn_Imag}).
As such, observing  these deviations could be used to directly probe steady-state Fano coherences.

More specifically, let the intensity of the emitted light in the absence of Fano coherence be $I^{(0)}$. This intensity  corresponds to the excited-state population in the absence of Fano coherence given by Eq.~(\ref{CanEqn_NoC}). From Eqs.~(\ref{CanEqn}) and (\ref{I_final2}), the relative difference of the emitted light intensity with and without steady-state Fano coherence is
\begin{equation}\label{DeltaI}
    \frac{\Delta I}{I^{0}} = \frac{\Delta \rho_{aa}}{\rho_{aa}^{(c)}} = \rho^{R}_{ab}
\end{equation} 
We see that in the coherent NESS induced by $x$-polarized incoherent driving of the Calcium V-system,  excited-state populations are suppressed compared to the canonical Boltzmann distribution. The intensity of the light emitted by the Calcium atoms  driven into the NESS is therefore diminished compared to the intensity that would be emitted by the atoms in coherence-free canonical thermal equilibrium (attainable by isotropic incoherent driving \cite{Tscherbul:14,Dodin:16,Dodin:18}). The relative difference in the fluorescence signals emitted by the atoms driven by $x$-polarized vs. isotropic incoherent light can therefore be interpreted as an experimental signature of   steady-state Fano coherences. 

\begin{figure}[t!]
\captionsetup{singlelinecheck = false, format= hang, justification=raggedright, font=footnotesize, labelsep=space}
\centering
\includegraphics[width=0.8\textwidth]{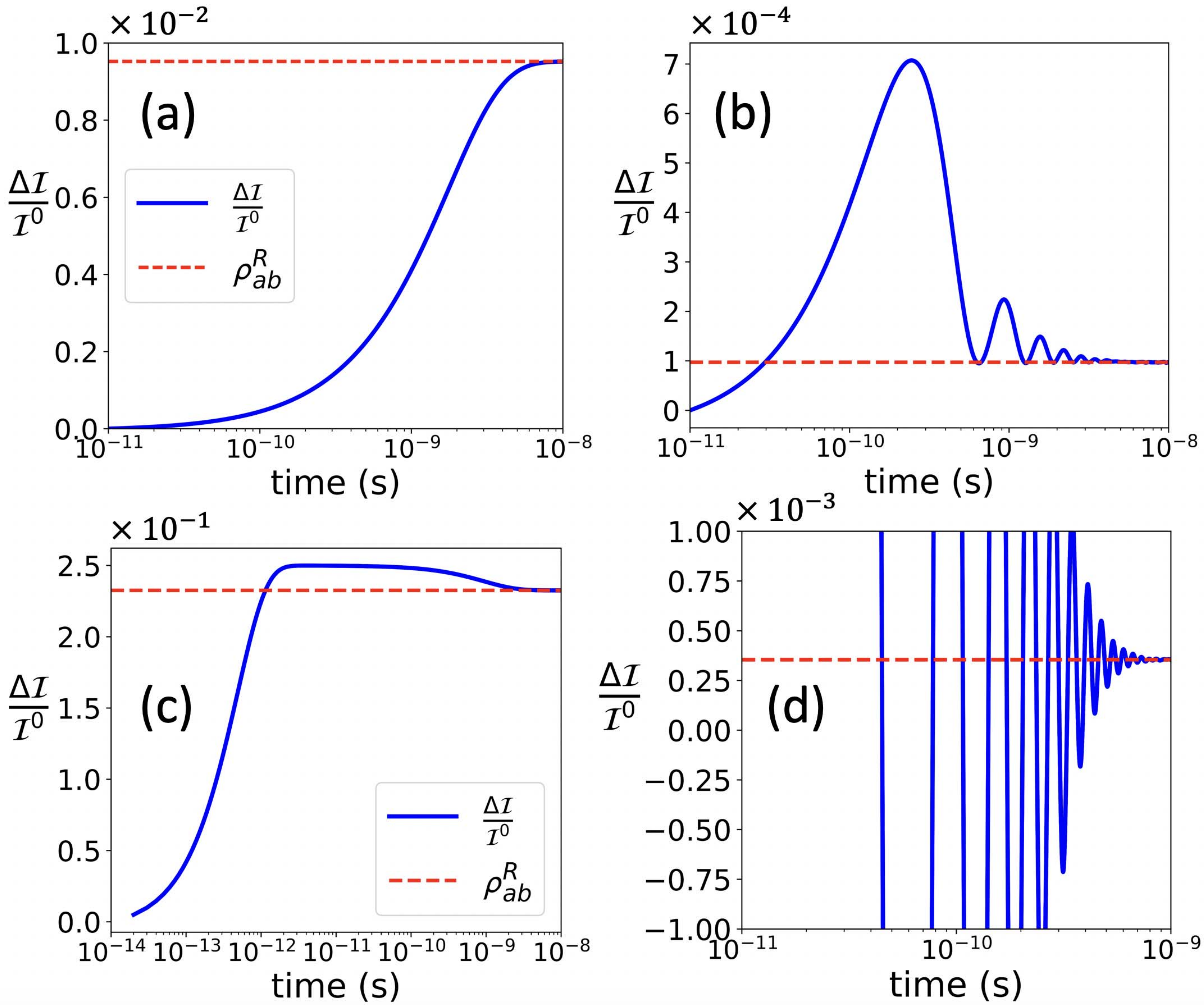}
\caption{(a) The ratio of fluorescence intensities calculated for Ca atoms driven by $x$-polarized vs. isotropic incoherent light  in different regimes: (a) $\bar{n}=10^{-2}$ and $\Delta/\gamma= 10^{-1}$; (b) $\bar{n}=10^{-2}$ and $\Delta/\gamma= 10$; (c) $\bar{n}=10^{3}$ and $\Delta/\gamma= 10$;  (d) $\bar{n}=10$ and $\Delta/\gamma= 10^{2}$ .}
\label{fig:5}
\end{figure}

Figure~\ref{fig:5} shows the experimentally relevant fluorescence difference signals calculated by numerical integration of the BR equations. We observe rich transient dynamics, after which the signals reach their non-equilibrium steady-state values consistent with Eq. (\ref{DeltaI}). The observable relative intensity difference remains fairly small ($<1\%$) in the weak-pumping limit  but grows significantly with the pumping intensity, reaching $23\%$ for $\bar{n}=10^3$ and $\Delta/\gamma=10$. As pointed out in Sec. IIIB above, it might be possible to further enhance the  steady-state Fano coherence (and hence the fluorescence difference signal) by inducing moderate decoherence between the excited states of the calcium atoms. This could be achieved experimentally by, {\it e.g}, adding magnetic field noise or isolating the atoms in a rare-gas host matrix \cite{Andrews:78,Upadhuay:19}.

%and Figure~(\ref{fig:5}a, b) show the relative intensity difference as a function of time in the weakly pumped V-system ($\bar{n} \ll 1$) in the over and underdamped regimes. Figure~(\ref{fig: Intensity_ratio}c, d) show the relative intensity difference as a function of time in the strongly pumped V-system ($\bar{n} \gg 1$) in the over and underdamped regimes. From these plots, we can conclude that the relative intensity difference signal is maximised in the overdamped regime.      

\section{Summary and conclusions}

We have presented closed-form analytical solutions to  the BR quantum master equations for a three-level V-system driven by polarized incoherent radiation, which reveal the presence of NESS featuring substantial  Fano coherences in the energy eigenstate basis. As pointed out in \cite{Agarwal:01}, the coherent NESS emerge as a result of an imbalance between polarized incoherent excitation and spontaneous emission. The imbalance occurs due to the directional nature of polarized incoherent pumping \cite{Dodin:18}, whereby only a single polarization mode of the radiation field interacts with the V-system \cite{Dodin:19}. As a result, polarized absorption rates are smaller by a factor of 4 that their isotropic counterparts \cite{Dodin:18} while the rates of spontaneous emission remain the same as in the case of unpolarized (isotropic) incoherent pumping, where both polarization modes participate in the emission process. The rates of isotropic incoherent pumping $r$ and spontaneous emission $\gamma$ are balanced, {\it i.e.} related by the equilibrium fluctuation-dissipation theorem $r=\bar{n}\gamma$, and thus no steady-state Fano coherences survive in the V-system driven by unpolarized incoherent light \cite{Agarwal:01}.

The imbalance  between absorption and spontaneous emission that leads to the emergence of steady-state Fano coherences can be thought of as originating from an additional, symmetry-breaking bath, leading to a formal analogy between the V-system driven by polarized incoherent light and a non-equilibrium transport problem involving a system of two qubits coupled to two thermal baths, one at $T=5800$~K responsible for incoherent solar excitation [see  Fig.~\ref{fig:1}(b)] and the other at zero temperature representing spontaneous emission.   We show that in this two-bath model,    the imaginary part of the coherence between the quasi-degenerate excited states in the V-system has a special significance: it is proportional to the non-equilibrium flux between the two-level systems coupled to hot and cold baths.

Our steady-state analysis demonstrates a remarkable relation between steady-state  Fano coherences  and the deviation of excited-state populations from thermal equilibrium: $\Delta \rho_{aa}/\rho^{(c)}_{aa} = \rho^{R}_{ab}$. This relation suggests that signatures of steady-state Fano coherences could be observed through measuring excited-state populations. We further extend our steady-state analysis to include environmental decoherence and relaxation effects, finding that  steady-state Fano coherence persists in the V-system illuminated by polarized incoherent radiation even in the presence of decoherence and relaxation effects.

While  steady-state Fano coherences are generally suppressed by environmental relaxation and decoherence (see Sec. IIIB), we find that the effect of  decoherence can be more subtle.
 In particular,  the real part of the steady-state Fano coherence can be enhanced by a moderate amount of excited-state decoherence, as shown in Fig.~\ref{fig:4}.   The magnitude of the steady-state Fano coherence is maximized when the decoherence rate is equal to the excited energy splitting minus the combined rates of radiative pumping and spontaneous decay ($\gamma_{d}=\Delta-r-\gamma$). We suggest that this counterintuitive enhancement of steady-state Fano coherence occurs due to a suppression of destructive interference of incoherent excitations generated at different times.
 
 % Further, we show that in a V-system driven by polarized incoherent radiation where, the phonon relaxation occurs through both the upward and downward transition between the ground and the excited state, the steady state coherence depends on the difference between the upward and downward transition rates, and it dissappears if the upward and downward transition rates are equal. \par 

Finally, we offer an improved method for the detection of steady-state Fano coherences using Calcium atoms irradiated by polarized incoherent light. Specifically, we propose to detect the signatures of Fano coherences by measuring the deviations of excited-state populations in the NESS from their equilibrium values (obtained by pumping the atoms with isotropic incoherent light). Unlike the scheme proposed in our previous work  \cite{Dodin:18}, this detection scenario  only relies on the total fluorescence intensity measurements, and requires neither spatially resolved fluorescence detection nor rapid turn-on of the incoherent light.
 %These deviations could be observed by detecting the deviation in the radiation emitted by the calcium atoms from the value expected in thermodynamic equilibrium.

\section*{Acknowledgements} P. B. acknowledges support from the U. S. Air Force office of Scientific Research  (AFOSR) under contract No. FA9550-17-1-0310. We thank Profs. Amar Vutha and Jonathan Weinstein  for stimulating discussions. 
%This work was supported by...

\newpage

\end{document}